\documentclass[preprint,12pt]{article}

  \usepackage{latexsym,bm,amsmath,amssymb,amsfonts}
  \usepackage{epsfig,graphics,graphicx}
  \usepackage{hyperref}
  \usepackage{slashed}
  \usepackage{cite}
  \usepackage{comment}
  \usepackage[latin1]{inputenc}
  \usepackage{mathrsfs}
\usepackage{mathcomp}

  \usepackage{lipsum}

\newcommand\blfootnote[1]{%
  \begingroup
  \renewcommand\thefootnote{}\footnote{#1}%
  \addtocounter{footnote}{-1}%
  \endgroup
}

  \newcommand{\beq}{\begin{eqnarray}}
  \newcommand{\eeq}{\end{eqnarray}}
  
 \def\simge{\mathrel{%
   \rlap{\raise 0.511ex \hbox{$>$}}{\lower 0.511ex \hbox{$\sim$}}}}
\def\simle{\mathrel{
   \rlap{\raise 0.511ex \hbox{$<$}}{\lower 0.511ex \hbox{$\sim$}}}}

\begin{document}

\begin{titlepage}

 \begin{flushright}
IPhT--T13/039\\
YITP--SB--13--08
 \end{flushright}

\bigskip
\bigskip

\centerline{\Large \bf Insane Anti--Membranes?}\bigskip

\medskip
\bigskip
\bigskip
\centerline{{\bf Gregory Giecold$^{\star}$, Francesco Orsi$^{\dagger}$\blfootnote{$^{\dagger}$Present address: Janelia Farm Research Campus, 19700 Helix Drive, Ashburn, VA 20147, U.S.A.} and Andrea Puhm$^{\ddagger}$}}
\centerline{{\bf }}
\bigskip
\centerline{$^{\star}$ C.N.~Yang Institute for Theoretical Physics,}
\centerline{State University of New York, Stony Brook, NY 11794--3840, U.S.A.}
\bigskip
\centerline{$^{\dagger \, \ddagger}$ Institut de Physique Th\'eorique, }
\centerline{CEA Saclay, CNRS--URA 2306, 91191 Gif--sur--Yvette, France}
\bigskip
\bigskip
\centerline{{\rm giecold@insti.physics.sunysb.edu, orsif@janelia.hhmi.org, andrea.puhm@cea.fr} }
\bigskip
\bigskip

\begin{abstract}

\noindent The backreactions of anti--branes on a variety of supergravity backgrounds have been shown in a recent series of papers to be riled by some unexplained flux singularities. All of the situations studied so far involve backgrounds with (close to) AdS--asymptotics. 
It is the purpose of this work to study the backreaction of anti--M2 branes on a background exhibiting a different UV behavior: The so--called $\mathbb{A}_8$ regular solution of eleven--dimensional supergravity that we consider has ``Taub--NUT type" asymptotics. As it turns out, some subleading infrared singularities are inevitable; they cannot be naturally ascribed to the anti--branes backreacting on this background.
Moreover, our configuration does not involve smeared branes. This lends further credence to the work of Bena et al.~\cite{Bena:2012vz} suggesting that the singularities encountered are in no way remnants of smearing that would wash away once brane polarization is taken into account.

\end{abstract}

\end{titlepage}

\tableofcontents
\newpage

\section{Introduction}
\setcounter{equation}{0}

Finding a mechanism to break supersymmetry in a controllable way is a challenge of major interest in string theory. This is important not only for the study of non--supersymmetric field theories in the context of the $AdS$/CFT correspondence but also for the construction and study of non-extremal black hole microstates.

One way to break supersymmetry is to put branes in backgrounds with charge dissolved in flux so that the charge of these branes is opposite with respect to that of the background. This mechanism exhibits quite interesting features: in numerous examples~\cite{Kachru:2002gs, DeWolfe:2004qx, Klebanov:2010qs, Bena:2011fc}, anti--branes in a probe approximation within some supergravity background have been found to give rise to metastable configurations. 
If the underlying supergravity theory has $AdS$--like asymptotics these configurations correspond to metastable states in the dual gauge theory and can be used to study the properties of non--supersymmetric gauge theories. Such was the purpose of~\cite{DeWolfe:2008zy} and~\cite{Bena:2011wh} where the expectation values of field theory operators in the putative metastable state are computed in terms of the asymptotic values of supergravity fields, and of the proposal for holographic mediation~\cite{Benini:2009ff, McGuirk:2011yg} where the visible sector gaugino mass is entirely determined by the holographic dual~\cite{McGuirk:2009xx, McGuirk:2009am}.

In order to determine if those brane putative metastable configurations are truly part of the same field theory as the supersymmetric states they are supposed to decay to, or, on the contrary if they actually are states of a different theory from the supersymmetric state, it is of great significance to go beyond the probe approximation and understand the backreaction of anti--branes embedded in some supergravity background.

In the completely different context of the fuzzball proposal~\cite{Mathur:2005zp, Bena:2007kg, Mathur:2008nj, Balasubramanian:2008da, Skenderis:2008qn, Mathur:2012dxa}, more complicated metastable brane configurations in a background with flat--space asymptotics are used to build microstates of non-extremal black holes~\cite{Bena:2011fc,Bena:2012zi}.
To see whether the physics of those metastable configurations as studied in \cite{Bena:2012zi} is the same after backreaction of the metastable supertubes, it is clearly important to understand the latter.
\\

The best studied example of anti--branes in a supergravity background is certainly that of anti--D3 branes in the Klebanov-Strassler (KS) warped deformed conifold~\cite{Klebanov:2000hb}. In a probe approximation\cite{Kachru:2002gs, DeWolfe:2004qx}, anti--D3 branes appear to give rise to metastable brane configurations that correspond to metastable states in the dual conformal field theory. It has been indeed advocated that such branes can tunnel to the dual supersymmetric minimum by annihilating their anti--brane charge against the positive charge of the background flux.

While the physics of these metastable configurations is quite appealing and an important ingredient e.g.~in string cosmology~\cite{Kachru:2003aw, Kachru:2003sx}, the recent body of work on the backreaction of anti--branes in Klebanov--Strassler~\cite{Bena:2009xk, Bena:2010ze, Bena:2011hz, Bena:2011wh, Dymarsky:2011pm} and other backgrounds in 11--dimensional supergravity~\cite{Bena:2010gs, Massai:2011vi} and type IIA supergravity~\cite{Giecold:2011gw} suggests that major problems arise which prevent a regular supergravity dual to metastability.

Indeed, for anti--D3 branes in KS, the fully backreacted supergravity solution exhibits singularities that are not directly sourced by any physical field\footnote{See~\cite{Massai:2012jn} for a very clear exposition.}. In the context of the $AdS$/CFT correspondence, it is generally admitted that only singularity--free solutions are dual to vacua of the gauge theory. 
Therefore, unless one finds a mechanism that resolves the ``unphysical'' singularities (and some proposals in this direction have been shown to fail)\footnote{See~\cite{Bena:2012tx, Bena:2012bk}, along with~\cite{Bena:2012vz}, which shows that brane polarization~\cite{Myers:1999ps} is highly unlikely to come to the rescue. Besides, references~\cite{Bena:2012ek, Bena:2013hr} establish that such singularities cannot be cloaked behind horizons, ruling out one of the criterion for a ``good singularity" put forth by Gubser~\cite{Gubser:2000nd}.}, it seems that one might be forced to discard those singular solutions.
\\

As already alluded to, another use of a supersymmetry--breaking process, from a different corner of string theory research, is related to supertubes in smooth ``bubbling" geometries~\cite{Bena:2011fc,Bena:2012zi}.
The background geometries have charge dissolved in flux on certain cycles (``bubbles") and represent microstates of extremal (supersymmetric) black holes. The supertube probes --- tubular brane configurations that have lower dimensional brane charge dissolved in worldvolume flux --- are placed in those geometries in such a way that either just one or both of the two electric charges of the supertube have an orientation opposite to that of the background charge.
The metastable configurations obtained this way break the supersymmetry (and extremality) of the background and are thus argued to correspond to microstates of non--extremal black holes. 
The study of non--extremal microstates is an important key in unlocking long--standing mysteries in black hole physics such as the information paradox. 
Given that the only two known fully--backreacted non--extremal microstates~\cite{Jejjala:2005yu,Giusto:2007tt,AlAlawi:2009qe,Bena:2009qv,Bobev:2009kn}  are very non-generic and their generalization is nowhere in sight, it is of great import to find the backreaction of these metastable supertubes in order to confirm the physics hinted by the probe analysis. 
This, however, is no easy task: it amounts to solving --- ideally analytically --- coupled partial differential equations in two variables. This would require either a generalization of the first--order formalism of~\cite{Borokhov:2002fm} for deforming around a supergravity background or a completely novel approach.

Given that unexplained, very likely non--physical, singularities appear in the backreaction of anti--D3 branes in warped deformed conifold and other such backgrounds, it is legitimate to ask whether backreacted metastable supertubes in bubbling geometries will suffer from the same problems?

It is assuredly not clear whether a naive extrapolation can be done for the arguments provided in~\cite{Bena:2012zi}. 
On the other hand, if it turns out that the backreacted solution does exhibit singularities that do not obviously stem from the metastable supertubes, this does not necessarily imply that the solution has to be discarded.
Indeed, in~\cite{Blaback:2012nf}, it was suggested that singularities in fluxes may arise from forcing time--independence on what is inherently a time-dependent process. As a matter of fact, non--extremal black holes radiate and so do their microstates. The appearance of a singularity may therefore imply that time--dependence of the full solution has to be taken into account and that Hawking radiation may, in fact, come from a perturbative decay~\footnote{We would like to thank T.~Van Riet for pointing out this possibility.}. 

It would be clearly important to investigate this further. But, first of all, one would need of course to know whether or not the backreaction of such black hole microstate geometries ``suffer'' from singularities or not.
\\

One of the purposes of this work is precisely to make progress in this direction. We want to see if the choice of $AdS$--like UV boundary conditions might somehow be directly responsible for the appearance of the unphysical singularities in the current series of papers on the backreaction of anti--branes~\cite{Bena:2009xk, Bena:2010ze, Bena:2011hz, Bena:2011wh, Bena:2010gs, Massai:2011vi, Giecold:2011gw}. 
All of the underlying background have $AdS$--like asymptotics and one could argue that the box--like nature of $AdS$--space acts as reflecting walls with a feedback loop on the fields that would prevent a singularity--free matching of the IR and UV conditions. It is a possibility that the fields that are sourced by the anti--branes in the IR go out to infinity but will be reflected back from the $AdS$ wall in the UV. Failure of these fields to match the way they initially looked like may then present itself in the form of singularities.

It is the aim of this work to study whether different UV boundary conditions for the BPS background on which the backreaction of anti--branes is then considered, can evade the appearance of singularities, apart from the one naturally sourced by the above--mentioned anti--branes.
\\

The second goal of the present paper is to further investigate the role of smearing of the anti--branes on some topological cycle, which is a key assumption of most papers on the backreaction of anti--branes. 

Arguments against smearing as the possible origin of the unphysical singularities have already been advanced in~\cite{Blaback:2010sj, Blaback:2011nz, Blaback:2011pn}, which consider anti--D6 branes localised in the transverse $S^3$. 
Moreover, in~\cite{Gautason:2013zw} it is shown that fully localised backreacting anti--D3 branes in a KS throat glued to a compactification manifold yields a singular energy density in the flux. These works thus suggest that smearing is not the source of singularities.

In the present work we give further credence to this argument. We carry out the first--order backreaction of anti--M2 branes on the $\mathbb{A}_8$ supergravity solution of~\cite{Cvetic:2001pga}, an asymptotically--flat space on which the anti--M2 branes are fully localised. The solution we obtain is still affected by such unphysical singularities.
\\ 

We follow the method of~\cite{Borokhov:2002fm} to obtain the full space of linearized deformations around this background. We have managed to describe this whole space in a fully analytical way. Our procedure is described in Section~\ref{Sect2} and the first three appendices.
Section~\ref{Sect3} contains our main results. Among the whole space of first--order deformation around the underlying BPS $\mathbb{A}_8$ regular supergravity solution, we obtain the solution describing the backreaction of anti--M2 branes on this background. This amounts to imposing the appropriate boundary conditions on the parameters mapping the space of linearized deformations.
As it happens, our solution is affected by some singularities which cannot be naturally seen as sourced by the anti--M2 branes 
located in the IR of the BPS background. As emphasized above, this raises questions on the fate of fully backreacted non--supersymmetric black hole microstate configurations.
Furthermore, our results add to the mounting evidence that the singularities encountered in all anti--brane backreactions considered to this date cannot possibly be attributed to the effect of smearing of those anti--branes, or to the $AdS$--like UV asymptotics of the underlying backgrounds.
\\

{\bf Note added:}\\
The week following the publication of our paper, a preprint~\cite{Cottrell:2013asa} appeared, which has significant overlap with our results. This provides another perspective on the analysis of linearized deformations around the $\mathbb{A}_8$ background.

\section{Deforming a supersymmetric background}\label{Sect2}
\setcounter{equation}{0}
We apply the method of Borokhov and Gubser~\cite{Borokhov:2002fm} that was also used in~\cite{Kuperstein:2003yt, Bena:2009xk, Bena:2010gs, Giecold:2011gw} to solve the eleven--dimensional supergravity equations perturbatively around a supersymmetric solution.
This is a general procedure that splits a set of $n$ second--order equations for $n$ fields $\phi^a$ that depend on a single radial variable $\tau$ to $2n$ first--order equations for $\phi^a$ and their conjugate variables $\xi_a$. 
 This method applies to supergravity problems whose symmetries only allow a dependence of the modes on a single, radial, variable. We refer to~\cite{Borokhov:2002fm} or one of the references mentioned above for a detailed description of this method.

\subsection{Ansatz for the perturbation}
We want to study supersymmetry--breaking deformations of a smooth solution to eleven--dimensional supergravity that corresponds to resolved M2 branes. This BPS solution is the so--called $\mathbb{A}_8$ background of Cvetic et al.~\cite{Cvetic:2001pga}. It has the peculiarity of not being asymptotically--AdS, not even by a whiff\footnote{See also~\cite{Hashimoto:2010bq}.}. In Section \ref{sec:0thorder} we provide the particular solutions to the field equations of the BPS $\mathbb{A}_8$ ansatz. In Appendix~\ref{AppAnsatz} we perform the reduction of this Ansatz to find a superpotential that is then plugged into the Borokhov--Gubser method~\cite{Borokhov:2002fm} for finding the first--order deformations 
\beq
\phi^{a}=\phi^{a}_0+\phi_1^{a}(X,Y) +\mathcal{O}(X^2,Y^2, X\, Y)\, \label{phipert}
\eeq
with a set $\{X_i,Y_i\}$, $i=1,...7$ of perturbation parameters around the regular supersymmetric $\mathbb{A}_8$ background. We will parametrize
\beq
\phi^{a} = \left( u, v, w, z, v_1, v_2, v_3 \right)\,
\eeq
in terms of the metric functions $(u,v,w)$ and gauge potential functions $(v_1,v_2,v_3)$ of 
the Ansatz encompassing the supersymmetric $\mathbb{A}_8$ solution to eleven--dimensional supergravity whose features we will discuss next.\\

For the metric, we take the Ansatz
\beq
ds_{11}^2 = H(r)^{-2/3} ~(-dt^2+dx_1^2+dx_2^2) + H(r)^{1/3} ~ds_8^2\, ,\label{eq:metricansatz}
\eeq
where the warp factor only depends on a radial coordinate of the 8--manifold with metric $ds_8^2$ which we take to have the same symmetries as the Spin(7) holonomy manifold $\mathbb{A}_8$ constructed in~\cite{Cvetic:2001pga}.
The Spin(7) holonomy garantees that the 11--dimensional metric and the four--form field strength admit at least one Killing spinor.
A convenient way to parametrize the warp factor and the metric on $\mathbb{A}_8$ is
\beq
H(r) = e^{3\, z(r)}\,,
\eeq
and
\beq
ds_8^2=\ell^2\, e^{- 2 w}\, dr^2 + e^{2 u}\, \left( D\mu^i \right)^2  + e^{2 w} \sigma^2 + e^{2 v}\, d\Omega_4^2\, ,\nonumber \label{ansatz}
\eeq
where $u,v,w$ are functions of the radial coordinate $r$ only and $\ell$ is a positive constant which we will later find convenient to set to unity. 

The 8-manifold is topologically an $S^3$ fiber over the $S^4$ with unit metric $d\Omega_4^2=\sum_{\alpha}(e^\alpha)^2$. The $S^3$ is itself an $S^1$ bundle with fiber $\varphi$ over $S^2$, with respective line elements $\sigma$ and $(D\mu^i)^2$ given by
\beq
\sigma=d\phi + \mathcal{A}\, \qquad \text{and} \qquad D\mu^i=d\mu^i + \varepsilon_{ijk} A^j \mu^k\,.
\eeq
The coordinates $\mu^i$ on the $S^2$ with $i=1,2,3$ are normalized in such a way that $\mu^i \mu^i=1$ and the $A^i$ are ${\bf su(2)}$ Yang-Mills instantons on the $S^4$. The full details of the $\mathbb{A}_8$ manifold can be found in~\cite{Cvetic:2001pga}.
Since the $S^3$ is ``squashed'' the metric allows for an asymptotic ``Taub-NUT type'' structure in which the $U(1)$ fiber $\varphi$ approaches a constant length while the radius of the $S^2$ grows linearly. \\

For the gauge potential we take the Ansatz
\beq \label{eq:gaugeansatz}
A_3= - K(r) ~dt \wedge dx_1 \wedge dx_2 + a_3\,, \qquad F_4 =dA_3
\eeq
where~\cite{Cvetic:2001pga}
\beq
a_3 = m ~\Big[ v_1\, \sigma \wedge X_2 + v_2\, \sigma \wedge J_2 + v_3\, Y_3 \Big]\,, \qquad G_4=da_3 \, ,\label{G4}
\eeq
gives rise to an anti-selfdual harmonic 4-form $G_4$.
The functions $K$ and $v_1$, $v_2,v_3$ depend on the radial coordinate $r$ only, and the 
2--foms $X_2$, $J_2$, along with the 3--form $Y_3$ on the $S^4$, are given by
\beq
X_2 \equiv \frac{1}{2}\, \epsilon_{ijk}\, \mu^i\, D\mu^j \wedge D\mu^k \, , \qquad J_2 \equiv \mu^i\, J^i \, , \qquad  Y_3 \equiv \epsilon_{ijk}\, \mu^i\, D\mu^j\, J^k \, .
\eeq
The $J^i$ are the field strengths of the Yang-Mills instanton potentials $A^i$:
\beq
J^i = dA^i + \frac{1}{2}\, \epsilon^{i}_{\, jk}\, A^j \wedge A^k\,.
\eeq

In this background the warp factor obeys
\beq
\square H = -\frac{1}{48} |G_4|^2\,.\label{warpeq}
\eeq

An algebraic constraint yields the radial derivative of the gauge potential $K$
\begin{align}\label{K prime solt}
\frac{d}{dr} \,K= - m^2\, \ell \, \Big[  4 v_1 v_3 + 4 v_2 v_3 -2 v_1 v_2 + v_2^2 \Big] e^{-(2u + 4v + 2w)} \, e^{- 6z}
\end{align}
In~\cite{Cvetic:2001pga}, the equations of motion for $u,v,w$ were obtained by varying the Lagrangian of the one dimensional sigma model while the equations for $v_1$, $v_2$ and $v_3$ were obtained by demanding that $G_4$ is harmonic\footnote{Alternatively, they can be derived from the flow equations stemming from the superpotential we have computed in Appendix \ref{AppAnsatz}}. 

If $G_4=0$ the above Ansatz describes ``bare'' M2 branes. The gauge potential \eqref{eq:gaugeansatz} is then given by
\beq
K (r) = \frac{1}{H(r)}\, ,\label{floating}
\eeq
which ensures that a probe M2 brane does not feel a force. The warp factor is a harmonic function of the 8--manifold; near $r=\ell$ it behaves as\footnote{Near $r=\ell$ the M2 brane has a coordinate singularity which corresponds to a horizon of topology $AdS_4 \times S^7$. The coordinate transformation $\tau \to \sqrt{r-\ell}$ amounts to shifting the horizon to $\tau=0$. The standard radial coordinate is thus given by $\tau$ for which the warp factor is sourced by an M2 brane with the familiar pole $\sim 1/\tau^6$.}
\beq
H \sim \frac{1}{(r-\ell)^3} + \mathcal{O}(r-1)^{-2}\,.
\eeq
While we will not be using the solution for bare M2 branes in this work we mention it here since it is of relevance to our later analysis of the boundary conditions imposed on the full--space of 1st--order deformations around the $\mathbb{A}_8$ background.
\\

The additional 4--form $G_4$ in \eqref{eq:gaugeansatz} can break the supersymmetry of the solution. The condition on the covariantly--constant spinor in $\mathbb{A}_8$ such that supersymmetry is preserved was given in~\cite{Cvetic:2001pga} and results in a linear relation between the functions $v_i$ of \ref{G4}. The regular $\mathbb{A}_8$ solution of \cite{Cvetic:2001pga} describing resolved M2 branes satisfies this relation and is thus supersymmetric. This solution, which we will summarize it in the next section, will be the starting point of our perturbation analysis.

\subsection{The $\mathbb{A}_8$ regular supergravity solution}\label{sec:0thorder}
The zeroth--order solutions $\phi^{a}_0= \left( u^0, v^0, w^0, z^0, v_1^0, v_2^0, v_3^0 \right)$ to the eleven--dimensional supergravity ansatz corresponding to resolved M2 branes with transverse $\mathbb{A}_8$ manifold are given by~\cite{Cvetic:2001pga}
\begin{align}\label{zeroth CGLP}
& e^{u^{0}} = \frac{1}{2}\, \sqrt{r + 3\, \ell}\, \sqrt{r - \ell}\, , \qquad e^{v^{0}} = \frac{1}{\sqrt{2}}\, \sqrt{r^2 - \ell^2} \, , \qquad e^{w^0} = \frac{\ell \, \sqrt{r + 3\, \ell}\, \sqrt{r - \ell}}{r+\ell} \, ,\nonumber\\
& v_1^{0}  = - \frac{\left( r - \ell \right)^2}{8\, \left( r + \ell \right)^2}\, , \qquad v_2^{0}  = \frac{\left( r - \ell \right)^2\, \left( r + 5\, \ell \right)}{8\, \left( r + \ell \right)\, \left( r + 3\, \ell \right)^2} \, , \qquad v_3^{0} = -\frac{\left( r - \ell \right)^2}{16\, \left( r + 3\, \ell \right)^2}\, .
\end{align}
The warp factor of the unperturbed solution is 
\beq
\label{0th z}
H_0 \equiv e^{3\, z^0} = \, 1 + m^2\, \frac{\left(3\, r^2 + 26\, \ell \, r + 63\, \ell^2 \right)}{20\, \ell \, \left( r + \ell \right)^2\, \left( r + 3\, \ell \right)^5}\, .
\eeq
For the gauge potential \eqref{K prime solt} we find
\beq
\frac{d}{dr}\, K_0 = - m^2\, \frac{\left(3\, r^3 + 33\, \ell\, r^2 + 121\, \ell^2\, r + 123\, \ell^3\right)}{4\, \ell (r+\ell)^3 (r+3\, \ell)^6}\, e^{-6z_0} \, ,
\eeq
which can easily seen to be related to the warp factor by
\beq
\frac{d\,K_0}{dr}= \frac{1}{H_0} \frac{d\, H_0}{dr}\,,
\eeq
as expected from~\eqref{floating}. For obvious practical reasons, from now on we will set $\ell = 1$.\\

As one can see, the solution of the uperturbed background corresponding to resolved M2 branes is \emph{smooth everywhere}.
It interpolates between eleven--dimensional Minkowski spacetime at small distances and $M_{2,1}\times S^1 \times \mathcal{M}_7$ at large distance where $\mathcal{M}_7$, an ${S}^3$ bundle over $S^4$, is an 7--manifold of $G_2$ holonomy. This is the asymptotic ``Taub--NUT type'' structure anticipated above.

\subsection{Force on a probe M2 brane}
The force on an M2 brane in a background with metric \eqref{eq:metricansatz} and gauge potential \eqref{eq:gaugeansatz} is determined by
\beq
V_\text{DBI} = H^{-1}\,, \qquad V_\text{WZ}= -K\,,
\eeq
so that
\beq
F= F_\text{DBI} + F_\text{WZ}= \frac{d}{dr} \Big(H^{-1} - K\Big)\,.
\eeq
For the zeroth--order solution with $K_0=H_0^{-1}$ the force $F^{(0)}=F_{\text{DBI}}^{(0)}+F_{\text{WZ}}^{(0)} $ on an M2 brane vanishes due to the Ansatz of~\eqref{floating}. For the family of linearized deformations represented by the set or perturbation parameters $\left\{X_i, Y_i\right\}$ in $\phi^a$ of \eqref{phipert} the force can be computed from the $2n$ linear first--order equations for the $\phi^a$ and their conjugate functions $\xi_a$. The result for the force $F^{(1)}=F_{\text{DBI}}^{(1)}+F_{\text{WZ}}^{(1)} $ on an M2 brane in this deformed supersymmetry--breaking background is
\beq 
F_{\text{DBI}}^{(1)} = - 3\, \left[ 3\, z^{0\, \prime}\, \phi_4 - \phi_4^{\prime} \right]\, e^{-3\, z^0} \, ,
\eeq
\begin{align}
F_{\text{WZ}}^{(1)} =&\, 2\, m^2\, \ell \, e^{-6\, z^0 - 2\, u^0 - 4\, v^0 - 2\, w^0}\, \Big[ \phi_5\, \left( - v_2^{0} + 2\, v_3^{0} \right) + \phi_6\, \left( - v_1^0 + v_2^0 + 2\, v_3^0 \right) \nonumber \\ & + 2\, \phi_7\, \left( v_1^0 + v_2^0 \right) - \left( \phi_1 + 2 \phi_2 + \phi_3 + 3\, \phi_4 \right)\, \left( -2\, v_1^0\, v_2^0 + v_2^{0\, 2} + 4\, v_1^0\, v_3^0 + 4\, v_2^0\, v_3^0 \right)\, \Big] .
\end{align}
Trading the radial derivative of $\phi_4$ through its equation of motion\footnote{This is a very lengthy expression which we therefore do not provide in this paper. Further details can be obtained upon request.} and making preemptive use of the analytic expression for $\tilde{\xi}_4$~\eqref{analyxi4}, it turns out that the force experienced by an M2 brane probing this background is
\beq\label{ForceM2}
F^{(1)} = \frac{X_4}{3}\, e^{-2\, u^0 -4\, v^0 - 2\, w^0} \, ,
\eeq
exactly, without any approximation or series expansion whatsoever.

In particular, this force depends on just a single mode. Such a result should by now be viewed as an expected end--product of the Borokhov--Gubser method~\cite{Borokhov:2002fm} on which we rely on to delineate the space of first--order deformations around a given supersymmetric background. Such a dependence of the force on a single mode is also observed in other setups~\cite{Bena:2009xk, Bena:2010gs, Giecold:2011gw}.

Note that this is a general result for the entire family of possible perturbation parameters $\left\{ X_i, Y_i \right\}$ entering \eqref{phipert}. In particular, this is independent of the boundary condition analysis of the modes that we are going to perform in the next section, in order to single out the particular solution describing the backreaction of anti--M2 branes on the $\mathbb{A}_8$ backgroud.

In the infrared, the force \eqref{ForceM2} goes as
\beq
F_{r} \sim \frac{X_4}{(r-1)^4} \, .
\eeq
In terms of the appropriate variable $\tau \sim \sqrt{r-1}$\footnote{The radial coordinate $\tau$ removes the coordinate singularity that is apparent in our expression for metric Ansatz written in terms of the radial variable $r$.} this becomes
\beq
F_{\tau} \sim \frac{dr}{d\tau}\, F_{r} \sim \tau\, X_4\, \frac{1}{\tau^8} = \frac{X_4}{\tau^7} \, ,
\eeq  
which is the behavior expected for the force on a probe M2 brane in the background of anti--M2 branes in 11--dimensional Minkowski space. 

\section{Obtaining the solution describing the backreaction of anti--M2 branes}\label{Sect3}
\setcounter{equation}{0}

We have managed to find fully analytical solutions for the modes $\phi^{a}$. Let us first remind the reader that they are written in terms of fourteen integration constants $\{X_i,Y_i\}$, $i=1,...,7$, of which seven ($X_i$) stem from the general solution to the system of auxiliary $\xi_{a}$ equations. Those integration constants are then fed into the system of $\phi^{a}$ equations through the latter's dependency on the $\xi_{a}$'s (see Appendix~\ref{AppPhi}). It is a non--trivial task to subsequently solve for this coupled system of $\phi^{a}$'s equations.
The analytic solutions which we managed to obtain involve the seven additional integration constants ($Y_i$).

The integration constants $X_i$ and $Y_i$ parametrize the full space of linear deformations around the regular BPS $\mathbb{A}_8$ eleven--dimensional supergravity solution~\cite{Cvetic:2001pga}. Our main interest is in identifying, among this space of general first--order deformations, the solution associated to the backreaction of anti--M2 branes sitting at the far infrared of the $\mathbb{A}_8$ background. This amounts to formulating conditions guided by the physics of this solution and imposing on the $X_i$'s and $Y_i$'s the ensuing relations.  

\subsection{IR boundary conditions}
\setcounter{equation}{0}

We proceed by first taking the infrared expansion around $r=1$ of the modes $\phi^a$. In the infrared sits a certain number $\bar{N}$ of anti--M2 branes. The supergravity fields should then behave as expected for M2 branes transverse to $\mathbb{A}_8$. This boils down to two physical conditions.
\\

\begin{itemize}
 \item 
 First of all, the warp factor should be proportional to $1/\tau^6$, as expected from solving the Laplace equation of a point charge in 8--dimensional space ($1/\tau^{8-2}$). In fact, the harmonic function on the 8--manifold of present interest appears in equation (55), Section 7.2 of~\cite{Cvetic:2001pga}. Its IR expansion involves polynomial pieces with decreasing power starting with $1/(r-1)^3$, along with an $\log(r-1)$ term. The leading behavior is just as expected, provided we keep in mind that the natural radial coordinate in the infrared is $\tau \sim \sqrt{r-1}$. 

From this first stipulation, it follows that the modes $\phi_{1,2,3}$ encoding the first--order deformations of the stretching functions to the transverse $\mathbb{A}_8$ geometry should not exhibit infrared divergences worse than $1/(r-1)^3$.

One might initially have thought that one should on top of that enforce the regularity of these modes, as is in fact done in the analysis of the backreaction of anti--D3 branes on the Klebanov--Strassler geometry~\cite{Bena:2011wh}. Yet, the stretching functions $u, v, w$ and their perturbations $\phi_{1,2,3}$ cannot be dissociated from the warp factor $z$ and its deformation, $\phi_4$. This is apparent from the form of the metric, where the relevant fields are not standalone $u$, $v$, $w$ but rather $z+2\, u$, $z+2\, v$ and $z+2\, w$:
\beq
& ds_{11}^2 = e^{-2 z}\, \eta_{\mu \nu}\, dx^{\mu}\, dx^{\nu} + e^{z}\, \left[ \ell^2\, e^{- 2 w}\, dr^2 + e^{2 u}\, \left( D\mu^i \right)^2 + e^{2 v}\, d\Omega_4 + e^{2 w} \sigma^2 \right]\, . \nonumber 
\eeq
Writing the BPS fields as $z_0$, $u_0$, etc.~the deformation of the metric is of the form
\begin{align}
& - 2\, \phi_4 e^{-2 z_0}\, \eta_{\mu \nu}\, dx^{\mu}\, dx^{\nu}  + e^{z_0}\, \Big[ \ell^2\, e^{- 2 w_0}\, (\phi_4 - 2\, \phi_3 ) dr^2 + \nonumber \\ &+e^{2 u_0}\, ( \phi_4 + 2\, \phi_1 ) \left( D\mu^i \right)^2 + e^{2 v_0}\, ( \phi_4 + 2\, \phi_2 ) d\Omega_4 + e^{2 w_0}\, ( \phi_4 + 2\, \phi_3 ) \sigma^2 \Big]\, . \nonumber
\end{align}
As such, if a leading $1/(r-1)^3$ divergence is allowed in $\phi_4$, it should be allowed in $\phi_{1,2,3}$ to boot, since there is no way of telling apart the origin of $1/(r-1)^3$ divergences in the above expression. 

One might wonder why the authors of~\cite{Bena:2011wh} decided to impose the regularity of the stretching functions as one of their IR boundary conditions. In fact, as we explain in Appendix~\ref{AppStretch}, these conditions need not be imposed: the same boundary conditions can be obtained by first focusing on the regularity of one of the modes associated to the 3--form fluxes.
\\ 

\item The second major physical condition is the requirement that the four--form flux $\sqrt{|G_4|^2}$ along the branes should go as $1/\tau^7$, as expected for anti--M2 brane sources. This means we have to suppress any term in the IR expansion of the modes $\phi_{5,6,7}$ (the perturbations to the four--form flux) that would otherwise yield a divergence worse than $1/\tau^{14}$ in the energy density $|G_4|^2 \sim (G_{4})_{i\, j\, k\, \ell}\, (G_{4})_{i'\, j'\, k'\, \ell'} g^{i\, i'}\, g^{j\, j'}\, g^{k\, k'}\, g^{\ell \, \ell'}$.
\\  
\end{itemize}

Let us now be more specific and derive the relations between the $X_i$'s and $Y_i$'s that are spawn by the two physical demands. This will successively fix particular integration constants in terms of the remaining ones.

The most divergent pieces of the IR expansions of $\phi_{1,2,3}$ are of order $1/(r-1)^4$. Banning such pieces fixes two of the $\left\{ X_i, Y_i \right\}$, which we take to be 
\beq
X_7 \, \, \, \, \, \& \, \, \, \, \, Y_3 \, .
\eeq

The series expansion of $\phi_4$ --- the perturbation to the warp factor --- then starts at order $1/(r-1)^3$. This is the behavior expected for the harmonic function of M2 branes. Our constraint is then to identify the term with  
\beq
\frac{Q_{IR}}{(r-1)^3} \, ,
\eeq
where $Q_{IR}$ measures the number of anti--M2 branes backreacting on their $\mathbb{A}_8$ background. This gives a condition on 
\beq
Y_7 \, 
\eeq
in terms of $Q_{IR}$ and the remaining integration constants, $X_{i}$, $i \neq 7$, and $Y_{i}$, $i \neq 3$.

The remaining IR conditions have to do with ensuring that the energy density $|G_4|^2$ behaves as expected from anti--M2 branes. Note that in previous setups~\cite{Bena:2009xk, Bena:2010gs, Giecold:2011gw} --- most importantly the backreaction of anti--D3 branes on the KS background --- this condition fails to be entirely consistent with the conditions imposed on the UV asymptotics of the deformation modes: unphysical singularities cannot be prevented. Similarly, we will see that such a hindrance arises for the backreaction of anti--M2 branes on $\mathbb{A}_8$ as well. 

We have to get rid of the $1/(r-1)^3$ and $1/(r-1)^2$ terms in the IR expansions of $\phi_{5,6,7}$. Otherwise the infrared behavior of $|G_4|^2$ would exhibit totally unacceptable singularities, that is to say $|G_4|^2$ would diverge more severely than the $1/(r-1)^7$ behavior that is characterizing anti--M2 branes.

Removing the $1/(r-1)^3$ terms in $\phi_{5,6,7}$ leads to a condition on 
\beq
X_6 \, .
\eeq
Culling the $1/(r-1)^2$ pieces in those modes spits out two constraints on 
\beq
X_5 \, \, \, \, \, \& \, \, \, \, \, X_4 \, .
\eeq
In particular, $X_4$ is entirely determined by the number of anti--M2 branes:
\beq
X_4 = -\frac{Q_{IR}}{3557323980000\, m^2} \, .
\eeq

Having determined four conditions on the $X_i$'s and two constraints on the $Y_i$'s from the physics associated to the M2--branes sitting in the infrared of their $\mathbb{A}_8$ background, we are about to shift our attention to the UV asymptotics of the backreaction of anti--M2's this background. 

\subsection{Matching the UV asymptotics}
\setcounter{equation}{0}

We want to make sure that the backreaction of anti--M2 branes on $\mathbb{A}_8$ preserves the original UV asymptotics of this BPS background. 

The UV expansions of $\phi_{1,2,3}$ each start at order $\mathcal{O}(r^0)$. As such, no condition needs to be imposed on these modes. 

We want to ensure that $\phi_5$, $\phi_6$ and $\phi_7$ are no more divergent than their BPS counterparts. Their UV expansions start at order $\mathcal{O}(r^3)$. As it happens, killing the $r^3$ pieces automatically guarantees that the sub--leading divergences of order $r^2$ and $r^1$ vanish as well. All in all, such conditions end up in a constraint on
\beq
Y_6 \, .
\eeq
The modes $\phi_{5,6,7}$ are now tamed.

As the next step, we look at the UV expansion of $\phi_4$. The term of order $r^5$ (and sub--leading divergent powers, it turns out) is culled by imposing a restriction on
\beq
Y_4 \, 
\eeq
in terms of $Q_{IR}$ and the integration constants that are still left unspecified thus far.

The IR and UV conditions exposed until now have given rise to four constraints on the $X_i$'s and four on the $Y_i$'s.

\subsection{Metric rescaling condition}
\setcounter{equation}{0}

Another integration constant can be gauged away by rescaling the three--dimensional Minkowski coordinates on the branes. The rescaling in $x_{\mu}$ is done by a constant shift in $\phi_4$. This condition produces another constraint on the $X_i$'s and $Y_i$'s, which we have decided to use to fix
\beq
Y_5 \, .
\eeq

\subsection{Zero--energy condition}
\setcounter{equation}{0}

There is one relation between the $X_i$, $i=1,...,7$ that has to be obeyed on the whole space of first--order deformations. This is the zero--energy condition on the kinetic energy terms and the potential of the reduction of eleven--dimensional supergravity on the $\mathbb{A}_8$ geometry:
\beq
T + V = 0 \, ,
\eeq
It is required to fully determine the reparametrization invariance of the radial coordinate~\cite{Gubser:2001eg}.
A condition on 
\beq
X_3
\eeq 
emanates, adding to a total of five constraints on $X_i$'s and five constraints on $Y_i$ as yet.

\subsection{M2 and M5 charges}
\setcounter{equation}{0}

We can define a running M2--charge by integrating the Hodge dual to the four--form flux $F_4$ at a fixed radial slice. Denoting by ${\cal M}_7$ the $\mathbb{S}^3$ bundle over $S^4$ characterizing the $\mathbb{A}_8$ geometry, let us introduce the M2 Maxwell charge, i.e.
\begin{align}
Q^{\text{Max}}_{M2}(r) & = \frac{1}{\left( 2 \, \pi \, \ell_p \right)^6}\, \int_{{\cal M}_7} *F_4(r) \nonumber\\ &
= \frac{\text{Vol}\left( {\cal M}_7 \right)}{\left( 2 \, \pi \, \ell_p \right)^6}\, \left[ v_1\, \left( -v_2 + 2\, v_3 \right) + v_2\, \left( - v_1 + v_2 + 2\, v_3 \right) + 2\, v_3 \, \left( v_1 + v_2 \right)\, \right] \, . 
\end{align}
Here, $\ell_p$ denotes the eleven--dimensional Planck length. In addition, the flux of $F_4$ threading the four-cycle that is present in the UV of the transverse $\mathbb{A}_8$ space is
\begin{align}
q &= - \frac{1}{\left( 2\, \pi\, \ell_p \right)^3} \int_{S^4} F_4(r_c) \nonumber\\ &
= \frac{2\, m\, \text{Vol}\left(S^4\right)}{\left( 2\, \pi\, \ell_p \right)^3}\, \left[ - v_1(r_c) + 2\, v_2(r_c)\right] \, .
\end{align}
We have introduced $r_c$ as the location of an UV boundary wall. This quantity is a measure of the fractional M5--charge that is present in the $\mathbb{A}_8$ background. 

Sending $r_c$ to infinity, the charges for the BPS $\mathbb{A}_8$ solution are found to be given by
\begin{align}
& Q^{\text{Max},\, (0)}_{M2}(r) = \frac{\text{Vol}\left( {\cal M}_7 \right)}{\left( 2 \, \pi \, \ell_p \right)^6}\, \frac{(r-1)^4\, \left(123+121\, r+33\, r^2+3\, r^3\right)}{64\, (1+r)^3\, (3+r)^4} \, , \nonumber\\ & q^{(0)} = \frac{3}{4}\, \frac{m\, \text{Vol}\left(S^4\right)}{\left( 2\, \pi\, \ell_p \right)^3} \, .
\end{align}
$Q^{\text{Max},\, (0)}_{M2}(r)$ interpolates between $\frac{35}{16384}\, \frac{\text{Vol}\left( {\cal M}_7 \right)}{\left( 2 \, \pi \, \ell_p \right)^6}\, (r-1)^4$ in the IR and $\frac{3}{64}\, \frac{\text{Vol}\left( {\cal M}_7 \right)}{\left( 2 \, \pi \, \ell_p \right)^6}$ in the UV.

The linear perturbations to the M2 and M5 charges are
\begin{align}
& Q^{\text{Max},(1)}_{M2}(r) = 2\, \frac{\text{Vol}\left( {\cal M}_7 \right)}{\left( 2 \, \pi \, \ell_p \right)^6}\, \left[ \phi_{5}\, \left( - v_2^0 + 2\, v_3^0 \right) + \phi_6\, \left( - v_1^0 + v_2^0 + 2\, v_3^0 \right) + 2\, \phi_7\, \left( v_1^0 + v_2^0 \right) \right] \, , \nonumber\\ 
&  q^{(1)} = \frac{2\, m\, \text{Vol}\left(S^4\right)}{\left( 2\, \pi\, \ell_p \right)^3}\, \left[ - \phi_5(r_c) + 2\, \phi_6(r_c) \right] \, .
\end{align}

We have to make sure that no extra M5--charge is introduced by the backreaction: $q^{(1)} = 0$. As it happens, $q^{(1)}$ is found to be directly proportional to the UV value of the perturbed M2 charge. Thus the UV M2 charge is left unchanged from its BPS value and we can use this condition to fix 
\beq
Y_2 \, .
\eeq

On the other hand, in the infrared, the perturbation to the M2-charge is determined by the following expansion\footnote{Taking into account the constraints on $X_i$ and $Y_i$ hitherto obtained.}:
\beq
\frac{Q_{IR}}{3336260466769920000}\, (r-1) + \mathcal{O}(r-1)^2 \, ,
\eeq
whence no extra boundary condition arises since this quantity vanishes when $r \to 1$. 
\\

Up to this point, we have determined $X_3$, $X_4$, $X_5$, $X_6$, $X_7$ and $Y_2$, $Y_3$, $Y_4$, $Y_5$, $Y_6$, $Y_7$ in terms of $Q_{IR}$, $X_1$, $X_2$ and $Y_1$. What other conditions can we possibly impose?

\subsection{Normalizability of the $\phi_a$ modes}
\setcounter{equation}{0}

The remaining integration constants $X_1$, $X_2$ and $Y_1$ are finally fixed by
imposing normalizability of the supergravity modes $\phi^a$. Indeed, some of them have pieces
going as $1/r$ in the UV. They integrate in the supergravity action to a divergent $\log(r)$, and
should therefore be eliminated. This is quite similar to the analysis of the boundary conditions associated to the backreaction of anti--D3 branes on the Klebanov--Strassler background~\cite{Bena:2009xk, Bena:2011wh, Dymarsky:2011pm}. 

This latter solution is close enough to being asymptotically--AdS that the usual holographic understanding of non--normalizable supergravity modes applies. Such modes are identified as irrelevant deformations on the field theory side; therefore, if one insists on guaranteeing that the metastable state and the supersymmetric state are states of the same field theory action, one must wipe off such non--normalizable modes.

Of course, the $\mathbb{A}_8$ background of present interest doesn't exhibit any AdS--asymptotics. Even though the holographic dictionary cannot be used, it is enough for our purposes to just view such normalizability conditions as related to a finite supergravity action.

When the dust settles, everything is determined by the parameter $Q_{IR}$, such that the infrared perturbation to the warp factor
\beq
\phi_{4} \sim \frac{Q_{IR}}{(r-1)^3} \, ,
\eeq 
and the full set of perturbation parameters $\{X_i,Y_i\}$ is specified in table \ref{Table:perturbationparams}.\\

\begin{table}[ht!]{\renewcommand{\arraystretch}{2}\renewcommand{\tabcolsep}{1cm}
\begin{center}{\normalsize
 \begin{tabular}{ll}
 $X_1 = -\frac{319\, Q_{IR}}{11383436736000000}$ & \hspace{0.5cm}$Y_1 = -\frac{11\, Q_{IR}\, \left[\, 90545597 + 83613600\, \log(2)^2\, \right]}{21249081907200000000}$ \\
 $X_2 = -\frac{103\, Q_{IR}}{3794478912000000}$  & \hspace{0.5cm}$Y_2 = -\frac{Q_{IR}\, \left[\, 90545597 + 83613600\, \log(2)^2\, \right]}{1328067619200000000}$\\
 $X_3 = -\frac{Q_{IR}}{758895782400000}$ & \hspace{0.5cm}$Y_3 = -\frac{13\, Q_{IR}\, \left[\, 90545597+83613600\, \log(2)^2\, \right]}{15936811430400000000}$\\
 $X_4=  -\frac{Q_{IR}}{3557323980000\, m^2}$ & \hspace{0.5cm}$Y_4 = -\frac{8743\, Q_{IR}\, \log(2)\, \left[\, 1092493 + 40\, \pi^2 + 320\, \log(2)^2\, \right]}{47366660947968000000}$\\
 $X_5 = \frac{29\, Q_{IR}}{2598393168000000}$ & \hspace{0.5cm}$Y_5 = \frac{Q_{IR}\, \left[\, 898989617 + 1219276800\, \log(2)^2\, \right]}{4406209624276992000000}$\\
 $X_6 = \frac{11\, Q_{IR}}{278399268000000}$ & \hspace{0.5cm}$Y_6 = \frac{Q_{IR}\, \left[\, 898989617 + 1219276800\, \log(2)^2\, \right]}{393411573596160000000}$\\
 $X_7 = \frac{47\, Q_{IR}}{649598292000000}$ & \hspace{0.5cm}$Y_7 = \frac{31\, Q_{IR}\, \left[\, 898989617 + 1219276800\, \log(2)^2\, \right]}{3934115735961600000000}$\\
\end{tabular}}
\end{center}\caption{List of all perturbation parameters.}\label{Table:perturbationparams}}
\end{table}

\section{Summary and conclusion}

Our ``executive summary" is that the force experienced by an M2--brane probing this background doesn't vanish and there is no IR singularity stemming from the energy density $|G_4|^2$, apart from the one associated to (anti--)M2 branes and subleading ones, whose physical significance is still a mystery. 

Indeed, similarly to the backreaction of anti--D3 branes on the KS background~\cite{Bena:2009xk, Bena:2011hz, Bena:2011wh}, along with other setups in 11--dimensional supergravity~\cite{Bena:2010gs, Massai:2011vi} and IIA~\cite{Giecold:2011gw}, there are nonetheless unaccounted--for singularities in the four--form flux. They could be considered unphysical in the sense that they pop up along directions not associated to the world--volume of anti--M2 branes. 

This work illustrates that such singularities are still present even when anti--branes are not smeared at the bottom of some topological cycle (as is the case in all the other backreacting solutions obtained to this date) and, just as significantly, when the underlying BPS background features flat--space UV--asymptotics. 
\\
\\

{\bf Acknowledgements}\\
We are thankful to Iosif Bena, Thomas Van Riet and Bert Vercnocke and especially Stefano Massai, for comments and discussions. Funding by the Research Foundation of Stony Brook University is appreciated. The work of A.P. is partly supported by the ANR grant 08--JCJC--0001--0. The work of G.G.~and A.P.~is sponsored partially by the ERC Starting Independent Researcher Grant 240210--String--QCD--BH. F.O.~is grateful to Janelia Farm for providing a pleasant working environment; much of this work was realized during his time at CEA Saclay, where he benefitted from fundings by the Consorzio Ferrare Ricerche (CFR) and in part from an ERC Starting Independent Researcher Grant 259133--ObservableString.
\\
\\

\appendix

\section{Reducing the Ansatz}\label{AppAnsatz}

In order to find the equations of motion for the fields entering the metric \eqref{eq:metricansatz} and the gauge potential \eqref{eq:gaugeansatz}, we reduce the bosonic part of the eleven--dimensional supergravity action
\beq\label{11sugra action}
S_{11} = \frac{1}{2 \, \kappa_{11}^2} \int d^{11}\, x\, \sqrt{\mid g \mid}\, \left[ R - \frac{1}{2} \mid F_4 \mid^2 \right] - \frac{1}{12 \, \kappa_{11}^2} \int A_3 \wedge F_4 \wedge F_4 \, .
\eeq
to a one--dimensional sigma model
\beq
\mathcal{S}_{11} = \frac{\text{Vol}\left(M_{1,2}\right)\, \text{Vol}\left( M_7 \right)}{2\, \kappa_{11}^2}\, \int dr\, \mathcal{L} \, ,
\eeq
where $\text{M}_{1,2}$ refers to the (2+1)--dimensional Minkowski space and ${\cal M}_{7}$ denotes the level surfaces of the 8--dimensional Spin(7) holonomy manifold. The equation of motion are then obtained by varying the Lagrangian $\mathcal{L} = T - V$ where
\begin{align}\label{kinetic}
T =&\, \frac{1}{\ell}\, e^{2 u + 4 v + 2 w} \, \left[ 2\, u^{\prime \, 2} + 16\, u^{\prime}\, v^{\prime} + 12\, v^{\prime \, 2} + 4\, u^{\prime}\, w^{\prime} + 8\, v^{\prime}\, w^{\prime} - \frac{9}{2}\, z^{\prime \, 2} \right] \nonumber\\ & -\frac{m^2}{2\, \ell}\, e^{-3 z + 2 u + 4 v + 2 w} \, \left[ \left( v_1^{\prime} \right)^2\, e^{-4 u - 2 w}+ 2\, \left( v_2^{\prime}\right)^2\, e^{-4 v - 2 w} + 4\, v_3^{\prime \, 2}\, e^{-2 u - 4 v} \right] \, ,
\end{align}
and (after eliminating the non-dynamical $K$ through its algebraic equation of motion)
\begin{align}\label{potential}
& V = \frac{\ell}{2}\, e^{- 2 u} \, \left[ 4\, e^{6 u} + 2\, e^{4 u + 2 w} - 24\, e^{4 u+ 2 v} - 4\, e^{2 u+ 4 v}+ e^{4 v + 2 w} \right] \nonumber\\ & + m^2\, \ell \, e^{-3 z + 2 u} \, \left[ \left(- v_1 + v_2 + 2\, v_3  \right)^2\, e^{-4u} + 2\, \left(-v_2 + 2\, v_3  \right)^2\, e^{-4 v} + 2\, \left(v_1+v_2\right)^2\, e^{-2u-2w}\right] \nonumber\\ & + \frac{1}{2}\, m^4\, \ell \, e^{-6 z - 2 u - 4 v - 2 w} \, \Big[ v_1  \, \left( -v_2 + 2\, v_3  \right) +  v_2 \, \left( - v_1 + v_2 + 2\, v_3 \right) + 2\, v_3\, \left( v_1 + v_2 \right) \Big]^2 \, . \nonumber\\
\end{align}
With the kinetic term
\beq
T = - \frac{1}{2}\, G_{ab}\, \frac{d{\phi^a}}{d r}\, 
\frac{d{\phi^b}}{d r}\,,
\eeq
where we denote the set of functions $\phi^{a}$ with $a = 1,\, ...,\, 7$ as
\beq\label{phi a set}
\phi^{a} = \left( u, v, w, z, v_1, v_2, v_3 \right) \, ,
\eeq 
we find that the superpotential
\begin{align}\label{The Superpotential}
W =&\, 2\, e^{w + 2 v}\, \left[ - 4\, e^{3 u} + 2\, e^{2 u + w} - 4\, e^{u+ 2 v} - e^{2 v + w} \right] \nonumber\\ & + 2\, m^2\, e^{-3 z}\, \Big[ v_1 \, \left( -v_2 + 2\, v_3 \right) + v_2 \, \left( - v_1 + v_2 + 2\, v_3 \right) + 2\, v_3\, \left( v_1 + v_2 \right) \Big]
\end{align}
accounts for all the terms in the potential~\eqref{potential} via
\beq\label{Vgr}
V = \frac{1}{8} \, G^{ab} \, \frac{\partial W}{\partial \phi^a}\, \frac{\partial W}{\partial \phi^b} \, .
\eeq

\section{$\xi^{i}$ equations}
\setcounter{equation}{0}

The first step of our analysis is to solve the system of equations for the
$\xi_i$'s.
To solve the $\xi^{i}$ equations in general, it is convenient to switch to the basis
\beq\label{New xi basis}
\tilde{\xi}_a = \left(\xi_1 - \xi_2 \, , \xi_2 - 2\, \xi_3 \, , \xi_1 - \xi_3 \, , \xi_4 \, , \xi_5 \, , \xi_6 \, , \xi_7 \right)\, .
\eeq
Then, in the order in which we will successively solve for the fields $\tilde{\xi}_a$, the equations are
\beq\label{xi4}
\tilde{\xi}_{4}^{\prime} = m^2 \, \ell \, e^{-2\, u^0 - 4\, v^0 - 2\, w^0  - 3\, z^0} \, \left[ 2\, v_1^0\, \left( v_2^0 - 2\, v_3^0 \right) - v_2^0\, \left( v_2^0 + 4\, v_3^0 \right) \right] \, \tilde{\xi}_4\, , 
\eeq
\beq\label{xi5}
\tilde{\xi}_{5}^{\prime} = - \frac{2}{3}\, m^2\, \ell \, e^{-2\, u^0 - 4\, v^0 - 2\, w^0 - 3\, z^0}\, \left( v_2^0 - 2\, v_3^0 \right)\, \tilde{\xi}_4 + \ell \, e^{-2 u^0}\, \tilde{\xi}_6 - \ell \, e^{-2 w^0}\, \tilde{\xi}_7 \, , 
\eeq
\begin{align}\label{xi6}
\tilde{\xi}_{6}^{\prime} = & - \frac{2}{3}\, m^2\, \ell \, e^{ - 2\, u^0 - 4\, v^0 - 2\, w^0 - 3\, z^0}\, \left( v_1^0 - v_2^0 - 2\, v_3^0 \right)\, \tilde{\xi}_4 \nonumber\\ & + 2\, \ell \, e^{2\, u^0 - 4\, v^0}\, \tilde{\xi}_5 - \ell \, e^{-2 u^0}\, \tilde{\xi}_6 - \ell \, e^{-2 w^0}\, \tilde{\xi}_7 \, , 
\end{align}
\beq\label{xi7}
\tilde{\xi}_{7}^{\prime} = \frac{2}{3}\, \ell e^{-2 \, u^0}\, \left[ 2\, m^2\, e^{- 4\, v^0 - 2\, w^0 - 3\, z^0}\, \left( v_1^0 + v_2^0 \right)\, \tilde{\xi}_4 - 6\, e^{4 \, u^0 - 4\, v^0}\, \tilde{\xi}_5 - 3\, \tilde{\xi}_6 \right] \, , 
\eeq
\begin{align}\label{xi1}
\tilde{\xi}_{1}^{\prime} =&\, \ell \, e^{-u^0-w^0}\, \left(-1+3\, e^{2\, u^0 - 2\, v^0}\right)\, \tilde{\xi}_1 + \ell \, \left( e^{-2\, v^0} - e^{-u^0-w^0} \right)\, \tilde{\xi}_2 \nonumber\\ & + \ell \, e^{-2\, u^0 - w^0}\, \left(2\, e^{u^0} + e^{w^0} \right)\, \tilde{\xi}_3 + 12\, \ell \, e^{2\, u^0 - 4\, v^0}\, \left( v_2^0 - 2\, v_3^0 \right)\, \tilde{\xi}_5 \nonumber\\ & - 2\, \ell \, e^{-2\, u^0}\, \left( v_1^0 - v_2^0 - 2\, v_3^0 \right)\, \tilde{\xi}_6 \nonumber\\ & + \frac{2}{3}\, m^2\, \ell \, e^{- 2\, u^0 - 4\, v^0 - 2\, w^0 - 3\, z^0}\, \left[ - 2\, v_1^0 \, \left( v_2^0 - 2\, v_3^0 \right) + v_2^0\, \left( v_2^0 + 4\, v_3^0 \right) \right]\, \tilde{\xi}_4 \, , 
\end{align}
\begin{align}\label{xi2}
\tilde{\xi}_{2}^{\prime} = &\, 2\, \ell \, e^{-u^0 - w^0}\, \tilde{\xi}_1 + \ell \, \left[ - e^{- 2 v^0} + 2\, e^{-u^0 - w^0} \right] \, \tilde{\xi}_2 - 4\, \ell \, e^{-u^0 - w^0} \, \tilde{\xi}_3 \nonumber\\ & - 8\, \ell \, e^{2\, u^0 - 4\, v^0}\, \left( v_2^0 - 2\, v_3^0 \right) \, \tilde{\xi}_5 - 4\, \ell\, e^{-2\, w^0}\, \left( v_1^0 + v_2^0 \right)\, \tilde{\xi}_7 \, , 
\end{align}
\begin{align}\label{xi3}
\tilde{\xi}_{3}^{\prime} =&\, 2\, \ell \, e^{u^0 - 2\, v^0 - w^0}\, \tilde{\xi}_1 + \ell \, e^{- 2\, u^0}\, \tilde{\xi}_3 + 4\, \ell \, e^{2\, u^0 - 4\, v^0}\, \left( v_2^0 - 2\, v_3^0 \right)\, \tilde{\xi}_5 \nonumber\\ & - 2\, \ell \, e^{- 2\, u^0}\, \left( v_1^0 - v_2^0 - 2\, v_3^0 \right)\, \tilde{\xi}_6 - 2\, \ell \, e^{- 2\, w^0}\, \left( v_1^0 + v_2^0 \right)\, \tilde{\xi}_7 \, .
\end{align}

The equation governing $\tilde{\xi}_{4}$ is immediately solved to
\beq\label{analyxi4}
\tilde{\xi}_{4}(r) = X_4 \, e^{3\, z^0(r)} \, .
\eeq

In fact, we have been successful in finding exact, analytic solutions for the entire system of coupled $\tilde{\xi}_{a}$ equations. They are quite lengthy and we refrain from publishing them here. They are available upon request from the authors.

\section{$\phi^{i}$ equations}\label{AppPhi}
\setcounter{equation}{0}

Once the $\xi^{i}$ equations have been solved analytically, one can insert these solutions in the set of $\phi^{a}$ equations, which we gather here. First of all, we introduce a helpful field redefinition\footnote{The inverse is $\phi^a = (\tilde{\phi}_1 \, , \tilde{\phi}_1 - \frac{1}{2}\, \tilde{\phi}_2 - \frac{1}{2}\, \tilde{\phi}_3 \, , - \tilde{\phi}_1 + \tilde{\phi}_3 \, , \tilde{\phi}_4 \, , \tilde{\phi}_5 \, , \tilde{\phi}_6 \, , \tilde{\phi}_7 )$}
\beq\label{phi redef}
\tilde{\phi}_{a} = (\phi_1 \, , \phi_1 - 2\, \phi_2 - \phi_3 \, , \phi_1 + \phi_3 \, , \phi_4 \, , \phi_5 \, , \phi_6 \, , \phi_7 )\, .
\eeq
Then, in the order in which we will successively solve for the fields $\tilde{\phi}_a$, the equations are
\begin{align}\label{phi1}
\tilde{\phi}_1^{\prime} = \frac{\ell}{12}\, e^{-2\, u^0 - 4\, v^0 -2 \, w^0}\, \Big[& \tilde{\xi}_1 + \tilde{\xi}_3 - 12\, e^{2\, v^0 + w^0}\, \big(e^{2\, v^0 + w^0}\, \tilde{\phi}_1 + e^{3\, u^0}\, \tilde{\phi}_2 + e^{u^0 + 2\, v^0}\, \tilde{\phi}_3 \big)\, \Big] \, ,
\end{align}
\begin{align}\label{phi2}
\tilde{\phi}_2^{\prime} = & \frac{\ell}{12} e^{-2\, u^0 - 4\, v^0 -2\, w^0}\, \Big[ \tilde{\xi}_1 + 4\, \big( \tilde{\xi}_3 - 3\, e^{2\, v^0 + w^0}\, \big( 2\, e^{2 v^0 + w^0}\, \tilde{\phi}_1 + 3\, e^{3\, u^0}\, \tilde{\phi}_2 + e^{u^0 + 2\, v^0}\, \tilde{\phi}_3 \big)\, \big)\, \Big] \, ,\nonumber\\
\end{align}
\begin{align}\label{phi3}
\tilde{\phi}_{3}^{\prime} = - \frac{\ell}{12}\, e^{-2\, u^0 -4\, v^0 -2\, w^0}\, \Big[& \tilde{\xi}_1 + 3\, \tilde{\xi}_2 - 2\, \Big( \tilde{\xi}_3 - 6\, e^{ u^0 + 2\, v^0 + w^0}\, \Big( e^{2\, u^0}\, \tilde{\phi}_2 \nonumber\\ & + e^{u^0 + w^0}\, \left( 2\, \tilde{\phi}_1 - \tilde{\phi}_2 - \tilde{\phi}_3 \right) + e^{2\, v^0}\, \tilde{\phi}_3 \Big)\, \Big)\, \Big] \, ,\nonumber\\
\end{align}
\begin{align}\label{phi5}
\tilde{\phi}_5^{\prime} = - \frac{\ell}{m^2}\, e^{2\, u^0 - 4\, v^0}\, \Big[& e^{3\, z^0}\, \tilde{\xi}_5 + 2\, m^2\, \Big( - 4 \, v_3^0\, \tilde{\phi}_1 + 2\, v_2^0\, \left( \tilde{\phi}_1 - 2\, \tilde{\phi}_2 \right) \nonumber\\ &
+ 8\, v_3^0\, \tilde{\phi}_2 + \tilde{\phi}_6 - 2\, \tilde{\phi}_7 \Big) \Big] \, ,
\end{align}
\begin{align}\label{phi6}
\tilde{\phi}_6^{\prime} = - \frac{\ell}{2\, m^2}\, e^{-2\, u^0}\, \Big[& e^{3\, z^0}\, \tilde{\xi}_6 + 2\, m^2\, \left( - 2\, v_1^0\, \tilde{\phi}_1 + 2\, v_2^0\, \tilde{\phi}_1 + 4\, v_3^0\, \tilde{\phi}_1 + \tilde{\phi}_5 - \tilde{\phi}_6 - 2\, \tilde{\phi}_7 \right) \Big] \, ,
\end{align}
\begin{align}\label{phi7}
\tilde{\phi}_7^{\prime} = - \frac{\ell}{4\, m^2}\, e^{-2\, w^0}\, \Big[& e^{3\, z^0}\, \tilde{\xi}_7 + 4\, m^2\, \left( 2\, v_1^0\, \tilde{\phi}_3 + 2\, v_2^0\, \tilde{\phi}_3 - \tilde{\phi}_5 - \tilde{\phi}_6 \right) \Big] \, ,
\end{align}
\begin{align}\label{phi4}
\tilde{\phi}_4^{\prime} =& - \frac{\ell}{9}\, e^{- 2\, u^0 - 4\, v^0 - 2\, w^0 - 3\, z^0}\, \Big[ e^{3\, z^0}\, \tilde{\xi}_4 + 3\, m^2\, \Big( - (v_2^0)^2\, \left( 2\, \tilde{\phi}_1 + 4\, \tilde{\phi}_2 + 2\, \tilde{\phi}_3 + 3\, \tilde{\phi}_4 \right) \nonumber\\ &
+ 4\, v_3^0\, \left( \tilde{\phi}_5 + \tilde{\phi}_6 \right) - 2\, v_2^0\, \left( v_3^0\, \left( 4\, \tilde{\phi}_1 + 8\, \tilde{\phi}_2 + 4\, \tilde{\phi}_3 + 6\, \tilde{\phi}_4 \right) + \tilde{\phi}_5 - \tilde{\phi}_6 - 2\, \tilde{\phi}_7 \right) \nonumber\\ &
- 2\, v_1^0\, \Big[ - v_2^0\, \left( 2\, \tilde{\phi}_1 + 4\, \tilde{\phi}_2 + 2\, \tilde{\phi}_3 + 3\, \tilde{\phi}_4 \right) \nonumber\\ &
+ v_3^0\, \left( 4\, \tilde{\phi}_1 + 8\, \tilde{\phi}_2 + 4\, \tilde{\phi}_3 + 6\, \tilde{\phi}_4 \right) + \tilde{\phi}_6 - 2\, \tilde{\phi}_7 \Big]\, \Big)\, \Big] \, ,\nonumber\\
\end{align}

It is worth noting that we have succeeded in finding an analytic solution to the system of $\tilde{\phi}_{1,2,3}$ equations. With such results in hand, we in turn successfully found exact solutions to the system of $\tilde{\phi}_{5,6,7}$ equations. Obtaining an analytic expression for $\tilde{\phi}_4$ is then a matter of a straightforward, if onerous integration. 

Those solutions are burdensome as such\footnote{A Mathematica file is available on demand.} but reduce to a more tractable form as soon as boundary conditions are imposed to pick a particular solution out of the full space of first--order deformations to the warped $\mathbb{A}_8$ background. Here, of course, our interest will be in setting the boundary conditions particularizing the backreaction of anti--M2 branes on the $\mathbb{A}_8$ background of~\cite{Cvetic:2001pga}.

\section{Conditions on the stretching functions}\label{AppStretch}
\setcounter{equation}{0}

As explained extensively in Section 3.1, our IR boundary conditions do not ask for the regularity of the perturbations $\phi_{1,2,3}$ to the stretching functions. This is seemingly in contrast to the analysis of the boundary conditions characterizing the backreaction of anti--D3 branes out of the full space of linear deformations to the warped deformed conifold.
\\ 

Actually, as we are now about to explain, it is not a crucial requirement to ask for the regularity of the stretching functions in~\cite{Bena:2011wh}. 

Instead, it is possible to rederive the boundary conditions associated to anti--D3's in KS, except that, this time, $1/\tau$
divergences in the modes $\phi_{1,2,3}$ are allowed a priori. These modes, which
are the perturbations to the stretching functions, are entangled with the
warp factor; as we claim, if a $1/\tau$ piece is allowed in $\phi_4$
(perturbation to the warp factor in~\cite{Bena:2011wh}), there is no good reason to kill the
$1/\tau$ terms in $\phi_{1,2,3}$.

By insisting on keeping a priori the $1/\tau$ pieces of $\phi_{1,2,3}$, one easily
obtains nearly the same boundary conditions as in~\cite{Bena:2011wh}. Interestingly, the relation
between $X_1$ and $X_6$ now comes from removing the $1/\tau$ divergence in
$\phi_7$. Formerly, in~\cite{Bena:2011wh}, the $1/\tau$ divergence of
$\phi_7$ would disappear automatically after imposing regularity of the
$\phi_{1,2,3}$ modes. We have turned the argument on its head.

In this revised analysis of the boundary conditions specifying the effects of anti--D3's, $Y_{2}^{IR}$ is left undetermined from what we have seen so far. In the boundary condition analysis of~\cite{Bena:2011wh}, it is set to zero by imposing that $\phi_2$ should not exhibit any divergent piece. Not anymore in our altered analysis. So how is it fixed? 

The answer is as follows: simply by requiring that the perturbations of $x-2p-A$ and of $-6p-x$
have the same coefficient in front of $1/\tau$. Remember that such
combinations are two identical ways of denoting the warp factor in the metric Ansatz of~\cite{Bena:2011wh}. This results in the condition that
\beq
\frac{12}{15}\, Y_2^{IR} - \frac{8}{10}\, Y_2^{IR} = 0 \, ,
\eeq
i.e.~$Y_2^{IR} = 0$. No such condition has to be imposed in the situation at hand involving our Ansatz for the perturbation around $\mathbb{A}_8$, given that our metric Ansatz features the warp factor in front of both the 3--dimensional Minkowski metric and as the $g_{rr}$ component of the metric.
\\

It is also important to stress the following distinction between our treatment of the stretching functions and the way they are handled when imposing the boundary conditions associated to anti--D3 branes in~\cite{Bena:2011wh}. In~\cite{Bena:2011wh}, the IR series of the stretching functions have terms going like $1/\tau$ but also like $\log(\tau)/\tau$.

As explained, one should keep these $1/\tau$ pieces, given that such a term is allowed in the mode describing the perturbation of the warp factor by anti--D3's. On the other hand, the warp factor associated to smeared branes goes like $1/\tau$ exactly; there are no subleading contributions, say, of the type $\log(\tau)/\tau$. For this very reason, such $\log(\tau)/\tau$ terms must be removed from the stretching functions by imposing apposite boundary conditions.

This should be distinguised from our present analysis of the boundary conditions prescribing the backreaction of anti--M2 branes on the $\mathbb{A}_8$ background, where subleading divergent terms are perfectly allowed in the stretching modes $\phi_{1,2,3}$. 

This has to do with the following observation: the harmonic function for M2 branes transverse to an $\mathbb{A}_8$ geometry is known to have an IR expansion involving an $1/(r-1)^3$ term, as well as $1/(r-1)^2$, $1/(r-1)$ and $\log(r-1)$ pieces, as can be seen in Section 7.2 of~\cite{Cvetic:2001pga}. 

For this reason and the fact that stretching functions and warp factors are intermingled and cannot be distinguished in the metric, such terms should not be removed when imposing boundary conditions on $\phi_{1,2,3}$. 



\begin{thebibliography}{30}

\bibitem{Kachru:2002gs}
  S.~Kachru, J.~Pearson and H.~L.~Verlinde,
  ``Brane/Flux Annihilation and the String Dual of a Non-Supersymmetric Field
  Theory,''
  JHEP {\bf 0206}, 021 (2002)
  \href{http://arxiv.org/abs/hep-th/0112197}{[arXiv:hep-th/0112197].}
  
\bibitem{DeWolfe:2004qx}
  O.~DeWolfe, S.~Kachru and H.~L.~Verlinde,
  ``The giant inflaton,''
  JHEP {\bf 0405}, 017 (2004)
  \href{http://arxiv.org/abs/hep-th/0403123}{[arXiv:hep-th/0403123].}
  
\bibitem{Klebanov:2010qs} 
  I.~R.~Klebanov and S.~S.~Pufu,
  ``M-Branes and Metastable States,''
  JHEP {\bf 1108}, 035 (2011)
  \href{http://arxiv.org/abs/1006.3587}{[arXiv:1006.3587 [hep-th]].}
  
\bibitem{Bena:2011fc} 
  I.~Bena, A.~Puhm and B.~Vercnocke,
  ``Metastable Supertubes and non-extremal Black Hole Microstates,''
  JHEP {\bf 1204}, 100 (2012)
  \href{http://arxiv.org/abs/1109.5180}{[arXiv:1109.5180 [hep-th]].}
  
\bibitem{DeWolfe:2008zy}
  O.~DeWolfe, S.~Kachru and M.~Mulligan,
  ``A Gravity Dual of Metastable Dynamical Supersymmetry Breaking,''
  Phys.\ Rev.\  D {\bf 77}, 065011 (2008)
  \href{http://arxiv.org/abs/0801.1520}{[arXiv:0801.1520 [hep-th]].}
  
\bibitem{Bena:2011wh} 
  I.~Bena, G.~Giecold, M.~Grana, N.~Halmagyi and S.~Massai,
  ``The backreaction of anti-D3 branes on the Klebanov-Strassler geometry,''
  \href{http://arxiv.org/abs/1106.6165}{arXiv:1106.6165 [hep-th].}
  
\bibitem{Benini:2009ff} 
  F.~Benini, A.~Dymarsky, S.~Franco, S.~Kachru, D.~Simic and H.~Verlinde,
  ``Holographic Gauge Mediation,''
  JHEP {\bf 0912}, 031 (2009)
  \href{http://arxiv.org/abs/0903.0619}{[arXiv:0903.0619 [hep-th]].}
  
\bibitem{McGuirk:2011yg} 
  P.~McGuirk,
  ``Hidden-sector current-current correlators in holographic gauge mediation,''
  Phys.\ Rev.\ D {\bf 85}, 045025 (2012)
  \href{http://arxiv.org/abs/1110.5075}{[arXiv:1110.5075 [hep-th]].}
  
\bibitem{McGuirk:2009xx}
  P.~McGuirk, G.~Shiu and Y.~Sumitomo,
  ``Non-supersymmetric infrared perturbations to the warped deformed
  conifold,''
  \href{http://arxiv.org/abs/0910.4581}{arXiv:0910.4581 [hep-th].}
  
\bibitem{McGuirk:2009am} 
  P.~McGuirk, G.~Shiu and Y.~Sumitomo,
  ``Holographic gauge mediation via strongly coupled messengers,''
  Phys.\ Rev.\ D {\bf 81}, 026005 (2010)
  \href{http://arxiv.org/abs/0911.0019}{[arXiv:0911.0019 [hep-th]].}
  
\bibitem{Mathur:2005zp} 
  S.~D.~Mathur,
  ``The Fuzzball proposal for black holes: An Elementary review,''
  Fortsch.\ Phys.\  {\bf 53}, 793 (2005)
  \href{http://arxiv.org/abs/hep-th/0502050}{[hep-th/0502050].}
  
\bibitem{Bena:2007kg} 
  I.~Bena and N.~P.~Warner,
  ``Black holes, black rings and their microstates,''
  Lect.\ Notes Phys.\  {\bf 755}, 1 (2008)
  \href{http://arxiv.org/abs/hep-th/0701216}{[hep-th/0701216].}
  
\bibitem{Mathur:2008nj} 
  S.~D.~Mathur,
  ``Fuzzballs and the information paradox: A Summary and conjectures,''
  \href{http://arxiv.org/abs/0810.4525}{arXiv:0810.4525 [hep-th].}
  
\bibitem{Balasubramanian:2008da} 
  V.~Balasubramanian, J.~de Boer, S.~El-Showk and I.~Messamah,
  ``Black Holes as Effective Geometries,''
  Class.\ Quant.\ Grav.\  {\bf 25}, 214004 (2008)
  \href{http://arxiv.org/abs/0811.0263}{[arXiv:0811.0263 [hep-th]].}
  
\bibitem{Skenderis:2008qn} 
  K.~Skenderis and M.~Taylor,
  ``The fuzzball proposal for black holes,''
  Phys.\ Rept.\  {\bf 467}, 117 (2008)
  \href{http://arxiv.org/abs/0804.0552}{[arXiv:0804.0552 [hep-th]].}
  
\bibitem{Mathur:2012dxa} 
  S.~D.~Mathur,
  ``Black holes and holography,''
  J.\ Phys.\ Conf.\ Ser.\  {\bf 405}, 012005 (2012)
  \href{http://arxiv.org/abs/1207.5431}{[arXiv:1207.5431 [hep-th]].}
  
\bibitem{Bena:2012zi}
  I.~Bena, A.~Puhm and B.~Vercnocke,
  ``Non-extremal Black Hole Microstates: Fuzzballs of Fire or Fuzzballs of Fuzz ?,''
  JHEP {\bf 1212} (2012) 014
  \href{http://arxiv.org/abs/1208.3468}{[arXiv:1208.3468 [hep-th]].}
  
\bibitem{Klebanov:2000hb}
  I.~R.~Klebanov and M.~J.~Strassler,
  ``Supergravity and a confining gauge theory: Duality cascades and
  chiSB-resolution of naked singularities,''
  JHEP {\bf 0008}, 052 (2000)
  \href{http://arxiv.org/abs/hep-th/0007191}{[arXiv:hep-th/0007191].}
  
\bibitem{Kachru:2003aw} 
  S.~Kachru, R.~Kallosh, A.~D.~Linde and S.~P.~Trivedi,
  ``De Sitter vacua in string theory,''
  Phys.\ Rev.\ D {\bf 68}, 046005 (2003)
  \href{http://arxiv.org/abs/hep-th/0301240}{[hep-th/0301240].}
  
\bibitem{Kachru:2003sx} 
  S.~Kachru, R.~Kallosh, A.~D.~Linde, J.~M.~Maldacena, L.~P.~McAllister and S.~P.~Trivedi,
  ``Towards inflation in string theory,''
  JCAP {\bf 0310}, 013 (2003)
  \href{http://arxiv.org/abs/hep-th/0308055}{[hep-th/0308055].}
  
\bibitem{Bena:2009xk}
  I.~Bena, M.~Gra\~na and N.~Halmagyi,
  ``On the Existence of Meta-stable Vacua in Klebanov-Strassler,''
  JHEP {\bf 1009}, 087 (2010)
  \href{http://arxiv.org/abs/0912.3519}{[arXiv:0912.3519 [hep-th]].}
  
\bibitem{Bena:2010ze} 
  I.~Bena, G.~Giecold, M.~Grana and N.~Halmagyi,
  ``On The Inflaton Potential From Antibranes in Warped Throats,''
  JHEP {\bf 1207}, 140 (2012)
  \href{http://arxiv.org/abs/1011.2626}{[arXiv:1011.2626 [hep-th]].}
  
\bibitem{Bena:2011hz} 
  I.~Bena, G.~Giecold, M.~Grana, N.~Halmagyi and S.~Massai,
  ``On Metastable Vacua and the Warped Deformed Conifold: Analytic Results,''
  Class.\ Quant.\ Grav.\  {\bf 30}, 015003 (2013)
  \href{http://arxiv.org/abs/1102.2403}{[arXiv:1102.2403 [hep-th]].}

\bibitem{Dymarsky:2011pm} 
  A.~Dymarsky,
  ``On gravity dual of a metastable vacuum in Klebanov-Strassler theory,''
  JHEP {\bf 1105}, 053 (2011)
  \href{http://arxiv.org/abs/1102.1734}{[arXiv:1102.1734 [hep-th]].}
  
\bibitem{Bena:2010gs} 
  I.~Bena, G.~Giecold and N.~Halmagyi,
  ``The Backreaction of Anti-M2 Branes on a Warped Stenzel Space,''
  JHEP {\bf 1104}, 120 (2011)
  \href{http://arxiv.org/abs/1011.2195}{[arXiv:1011.2195 [hep-th]].}
  
\bibitem{Massai:2011vi} 
  S.~Massai,
  ``Metastable Vacua and the Backreacted Stenzel Geometry,''
  JHEP {\bf 1206}, 059 (2012)
  \href{http://arxiv.org/abs/1110.2513}{[arXiv:1110.2513 [hep-th]].}

\bibitem{Giecold:2011gw} 
  G.~Giecold, E.~Goi and F.~Orsi,
  ``Assessing a candidate IIA dual to metastable supersymmetry-breaking,''
  JHEP {\bf 1202}, 019 (2012)
  \href{http://arxiv.org/abs/1108.1789}{[arXiv:1108.1789 [hep-th]].}
  
\bibitem{Massai:2012jn} 
  S.~Massai,
  ``A Comment on anti-brane singularities in warped throats,''
  \href{http://arxiv.org/abs/1202.3789}{arXiv:1202.3789 [hep-th].}
  
\bibitem{Bena:2012tx} 
  I.~Bena, D.~Junghans, S.~Kuperstein, T.~Van Riet, T.~Wrase and M.~Zagermann,
  ``Persistent anti-brane singularities,''
  JHEP {\bf 1210}, 078 (2012)
  \href{http://arxiv.org/abs/1205.1798}{[arXiv:1205.1798 [hep-th]].}

\bibitem{Bena:2012bk} 
  I.~Bena, M.~Grana, S.~Kuperstein and S.~Massai,
  ``Anti-D3's - Singular to the Bitter End,''
  \href{http://arxiv.org/abs/1206.6369}{arXiv:1206.6369 [hep-th].}
  
\bibitem{Bena:2012vz} 
  I.~Bena, M.~Grana, S.~Kuperstein and S.~Massai,
  ``Polchinski-Strassler does not uplift Klebanov-Strassler,''
  \href{http://arxiv.org/abs/1212.4828}{arXiv:1212.4828 [hep-th].}
  
\bibitem{Myers:1999ps}
  R.~C.~Myers,
  ``Dielectric-branes,''
  JHEP {\bf 9912}, 022 (1999)
  \href{http://arxiv.org/abs/hep-th/9910053}{[arXiv:hep-th/9910053].}
  
\bibitem{Bena:2012ek} 
  I.~Bena, A.~Buchel and O.~J.~C.~Dias,
  ``Horizons cannot save the Landscape,''
  \href{http://arxiv.org/abs/1212.5162}{arXiv:1212.5162 [hep-th].}
  
\bibitem{Bena:2013hr} 
  I.~Bena, J.~Blaback, U.~H.~Danielsson and T.~Van Riet,
  ``Antibranes don't go black,''
  \href{http://arxiv.org/abs/1301.7071}{arXiv:1301.7071 [hep-th].}
  
\bibitem{Gubser:2000nd} 
  S.~S.~Gubser,
  ``Curvature singularities: The Good, the bad, and the naked,''
  Adv.\ Theor.\ Math.\ Phys.\  {\bf 4}, 679 (2000)
  \href{http://arxiv.org/abs/hep-th/0002160}{[hep-th/0002160].}
  
\bibitem{Jejjala:2005yu}
  V.~Jejjala, O.~Madden, S.~F.~Ross and G.~Titchener,
  ``Non-supersymmetric smooth geometries and D1-D5-P bound states,''
  Phys.\ Rev.\ D {\bf 71} (2005) 124030
  \href{http://arxiv.org/abs/hep-th/0504181}{[hep-th/0504181].}
  
\bibitem{Giusto:2007tt}
  S.~Giusto, S.~F.~Ross and A.~Saxena,
  ``Non-supersymmetric microstates of the D1-D5-KK system,''
  JHEP {\bf 0712} (2007) 065
  \href{http://arxiv.org/abs/0708.3845}{[arXiv:0708.3845 [hep-th]].}
  
\bibitem{AlAlawi:2009qe}
  J.~H.~Al-Alawi and S.~F.~Ross,
  ``Spectral Flow of the Non-Supersymmetric Microstates of the D1-D5-KK System,''
  JHEP {\bf 0910} (2009) 082
  \href{http://arxiv.org/abs/0908.0417}{[arXiv:0908.0417 [hep-th]].}
  
\bibitem{Bena:2009qv}
  I.~Bena, S.~Giusto, C.~Ruef and N.~P.~Warner,
  ``A (Running) Bolt for New Reasons,''
  JHEP {\bf 0911} (2009) 089
  \href{http://arxiv.org/abs/0909.2559}{[arXiv:0909.2559 [hep-th]].}
  
\bibitem{Bobev:2009kn}
  N.~Bobev and C.~Ruef,
  ``The Nuts and Bolts of Einstein-Maxwell Solutions,''
  JHEP {\bf 1001} (2010) 124
  \href{http://arxiv.org/abs/0912.0010}{[arXiv:0912.0010 [hep-th]].}
  
\bibitem{Borokhov:2002fm}
  V.~Borokhov and S.~S.~Gubser,
  ``Non-supersymmetric deformations of the dual of a confining gauge  theory,''
  JHEP {\bf 0305}, 034 (2003)
  \href{http://arxiv.org/abs/hep-th/0206098}{[arXiv:hep-th/0206098].}
  
\bibitem{Blaback:2012nf} 
  J.~Blaback, U.~H.~Danielsson and T.~Van Riet,
  ``Resolving anti-brane singularities through time-dependence,''
  JHEP {\bf 1302}, 061 (2013)
  \href{http://arxiv.org/abs/1202.1132}{[arXiv:1202.1132 [hep-th]].}
  
\bibitem{Blaback:2010sj}
  J.~Blab\"ack, U.~H.~Danielsson, D.~Junghans, T.~Van Riet, T.~Wrase and M.~Zagermann,
  ``Smeared versus localised sources in flux compactifications,''
  \href{http://arxiv.org/abs/1009.1877}{arXiv:1009.1877 [hep-th]}.
  
\bibitem{Blaback:2011nz} 
  J.~Blaback, U.~H.~Danielsson, D.~Junghans, T.~Van Riet, T.~Wrase and M.~Zagermann,
  ``The problematic backreaction of SUSY-breaking branes,''
  JHEP {\bf 1108}, 105 (2011)
  \href{http://arxiv.org/abs/1105.4879}{[arXiv:1105.4879 [hep-th]].}
  
\bibitem{Blaback:2011pn} 
  J.~Blaback, U.~H.~Danielsson, D.~Junghans, T.~Van Riet, T.~Wrase and M.~Zagermann,
  ``(Anti-)Brane backreaction beyond perturbation theory,''
  JHEP {\bf 1202}, 025 (2012)
  \href{http://arxiv.org/abs/1111.2605}{[arXiv:1111.2605 [hep-th]].}
  
\bibitem{Gautason:2013zw} 
  F.~F.~Gautason, D.~Junghans and M.~Zagermann,
  ``Cosmological Constant, Near Brane Behavior and Singularities,''
  \href{http://arxiv.org/abs/1301.5647}{arXiv:1301.5647 [hep-th].}
  
\bibitem{Cvetic:2001pga}
  M.~Cveti$\check{c}$, G.~W.~Gibbons, H.~L\"u and C.~N.~Pope,
  ``New complete non-compact Spin(7) manifolds,''
  Nucl.\ Phys.\  B {\bf 620}, 29 (2002)
  \href{http://arxiv.org/abs/hep-th/0103155}{[arXiv:hep-th/0103155].}
  
\bibitem{Kuperstein:2003yt}
  S.~Kuperstein and J.~Sonnenschein,
  ``Analytic non-supersymmetric background dual of a confining gauge theory
  and the corresponding plane wave theory of hadrons,''
  JHEP {\bf 0402}, 015 (2004)
  \href{http://arxiv.org/abs/hep-th/0309011}{[arXiv:hep-th/0309011].}
  
\bibitem{Gubser:2001eg}
  S.~S.~Gubser, A.~A.~Tseytlin and M.~S.~Volkov,
  ``NonAbelian 4-d black holes, wrapped five-branes, and their dual descriptions,''
  JHEP {\bf 0109} (2001) 017
  \href{http://arxiv.org/abs/hep-th/0108205}{[arXiv:hep-th/0108205].}
  
\bibitem{Hashimoto:2010bq}
  A.~Hashimoto, S.~Hirano and P.~Ouyang,
  ``Branes and fluxes in special holonomy manifolds and cascading field
  theories,''
  \href{http://arxiv.org/abs/1004.0903}{arXiv:1004.0903 [hep-th].}
  
\bibitem{Cottrell:2013asa} 
  W.~Cottrell, J.~Gaillard and A.~Hashimoto,
  ``Gravity dual of dynamically broken supersymmetry,''
  \href{http://arixv.org/abs/1303.2634}{arXiv:1303.2634 [hep-th].}
 
 
\end{thebibliography}
\end{document}